\documentclass[%
 aip,
 amsmath,amssymb,
 reprint,%
]{revtex4-1}

\usepackage{graphicx}
\usepackage{dcolumn}
\usepackage{bm}

\usepackage[utf8]{inputenc}
\usepackage[T1]{fontenc}
\usepackage{mathptmx}
\usepackage{etoolbox}
\usepackage{comment}

\usepackage{booktabs}

\makeatletter
\def\@email#1#2{%
 \endgroup
 \patchcmd{\titleblock@produce}
  {\frontmatter@RRAPformat}
  {\frontmatter@RRAPformat{\produce@RRAP{*#1\href{mailto:#2}{#2}}}\frontmatter@RRAPformat}
  {}{}
}%
\makeatother
\begin{document}

\preprint{AIP/123-QED}

\title{
Raman mass-spectrometry via oscillatory motion in deep pulsed optical lattices
}
\author{A. Gerakis}
 \email{alexandros.gerakis@list.lu.}
\affiliation{ 
Materials Research \& Technology Department, Luxembourg Institute of Science \& Technology, L-$4422$, Belvaux, Luxembourg\\
}%
\author{M. N. Shneider}%
\affiliation{ 
Department of Mechanical \& Aerospace Engineering, Princeton University, $08544$, Princeton, NJ, USA
}%

\author{P. F. Barker}
\affiliation{%
Department of Physics \& Astronomy, University College London, WC$1$E $6$BT, London, UK
}%

\date{\today}

\begin{abstract}
We describe a new optical diagnostic for determining the composition of gases by measuring the motion of atoms and molecules trapped within very deep optical lattices. This non-resonant method is analogous to conventional Raman scattering, except that the observed spectral features relate to the oscillatory center-of-mass motion of each species within the lattice, determined uniquely by their respective polarizability-to-mass ratio. Depending on the density of the probed sample, detection occurs either via optical scattering at the high end or via non-resonant ionization at the lower end. We show that such a technique is complementary to conventional mass spectrometry techniques and envision its implementation in existing instruments.
\end{abstract}

\maketitle

Mass spectrometry describes a general family of established and powerful analytical techniques that are used widely to determine and quantify the chemical composition of a wide range of organic and inorganic materials. It is commonly used to identify unknown compounds within a sample, and to elucidate the structure and chemical properties of biomolecules. Depending on the selected mass spectrometry technique, the user can additionally obtain spatially resolved information (in 1D, 2D and 3D).
However, as smaller feature sizes in nanotechnology require higher spatial resolution, new methods with higher 
detection limits and spatial accuracies are needed. 

Traditional mass spectrometry techniques rely on the ionization of the sample's particles and subsequent filtering according to the mass-to-charge ratio. 
Although undoubtedly successful, this approach sometimes has inherent limitations 
in the achievable resolution and quantitive analysis~\cite{doi:10.1098/rsta.2015.0366,Wirtz_2015}. For example,   
secondary ion mass spectrometry (SIMS), which is an extremely powerful tool for surface and thin film characterization, has excellent detection limits and high dynamic range with nanoscale resolution~\cite{Audinot_2021, PILLATSCH20191, 10.3389/fnbeh.2020.00124}. However, it experiences relatively low and non-uniform ionization ratios for the sputtered particles from the sample, rendering quantitative analysis challenging with no clear path to further improvement of the spatial resolution. Similar challenges are faced by other major types of mass-spectrometry, such as matrix assisted laser desorption ionization-time of flight (MALDI-TOF)~\cite{10.3389/fmicb.2015.00791,aichler2015maldi} and the  Orbitrap~\cite{doi:10.1021/ac4001223, doi:10.1146/annurev-anchem-071114-040325, https://doi.org/10.1002/mas.20186}. 

To address some of these challenges, the use of laser probes has been studied 
to enhance the post-ionization of the sputtered material by utilizing resonant or non-resonant radiation. Limited enhancement has been demonstrated in the former case due to the 
low ionization cross-sections while in the latter case due to the limited number of species that can be simultaneously ionized~\cite{KING2003244,Wise1995,Kollmer2003,He1999,Mibuka2008,Ebata2013}.

In this Letter, we propose a novel scheme for mass spectrometry based on the polarizability-to-mass ratio, which relies on the coherent oscillation of trapped species within high-intensity optical lattices\cite{Bishop_2010,PhysRevLett.93.243004, maher2012laser}. This phenomenon has been observed when ultra cold atomic species in the $\mu$K regime are trapped within shallow optical lattices (T$\sim$mK). However, it has not  been used for the hot (T>$300$ K) 
species considered here, as usually the light intensity is too low. Our scheme overcomes the limitations currently faced by 
the state-of-the-art mass spectrometry techniques, by allowing for the non-resonant detection of all species of the periodic table and their isotopes, while furthermore resolving mass interferences among different elements/compounds. In addition, this scheme interacts equally with all species in the measurement volume, hence enabling quantitative measurements. Importantly, it is also applicable to atomic and molecular species, and their compounds, in their ground or excited states. It is complimentary to the already established mass spectrometry techniques by coupling them with existing laser sources. Depending on the measurement conditions, we describe two different ways of species detection: optical detection for high density samples and detection for low density samples using a final non-resonant ionization stage.







\begin{figure}[!h]
\includegraphics[width=0.99\columnwidth]{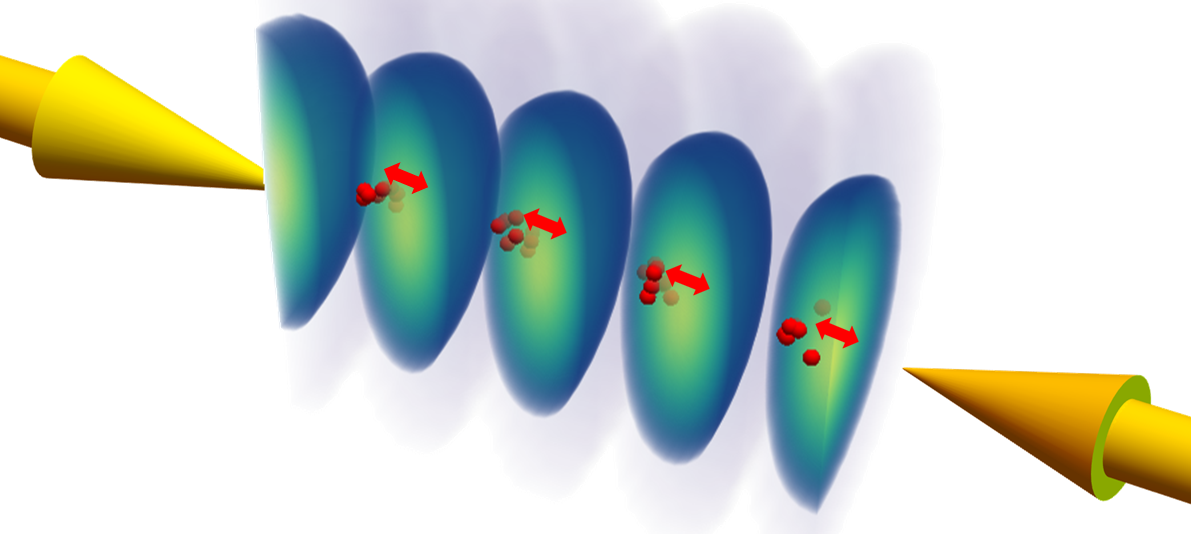}
\caption{\label{fig:one} An optical lattice created by two counterpropagating Gaussian optical beams (indicated by the yellow arrows). The optical lattice created is shown using a contour plot of the light intensity. Only one half of the lattice is shown to illustrate the variation of light intensity. Here yellow is high intensity while dark blue is low intensity. The red spheres indicate an ensemble of neutral particles that at this location are attracted towards the high intensity part of the lattice following near harmonic motion within the induced optical dipole potential of the lattice field.    }
\end{figure}

Our approach to species specific detection relies on the non-resonant optical dipole force (ODF). For fields that are far-
detuned from dipole allowed resonances, the optical dipole force is proportional to the DC polarizability $\alpha$ of the atomic or molecular species and the gradient in the optical field $E$ averaged over an optical cycle. The equation of motion of a species in any direction $j$ is given by $\frac{d^2x_j}{dt^2}=\frac{1}{4}\frac{\alpha}{m} \nabla E_j^2$. The center-of-mass motion in this field is proportional to the particle's polarizability-to-mass ratio, thus offering a mechanism for dispersion in analogy with the charge-to-mass ratio utilized typically in conventional mass spectrometry. As the optical dipole force is generally weak, the field strength and gradient need to be maximised to enhance the light-particle interaction. This can be achieved effectively in a optical lattice formed by the interaction of the particles to be detected with the optical interference pattern formed between two counter-propagating light fields, as shown in Fig.~\ref{fig:one}. This geometry provides a large gradient in the optical field while the use of pulsed lasers allows for the use of large field strengths, effectively limited only by ionization. Typical laser intensities operating below the ionization threshold would be on the order of $10^{15}$ W m$^{-2}$ which can easily be delivered by standard commercial nanosecond pulsed lasers.  

Neutral particles can be trapped in the resulting optical lattice due to the ODF for the duration of the pulse even for room temperature gases. For relatively long ($>10$ ns) duration pulses the particles perform oscillatory motion around the antinode of the interference pattern. The frequency of this oscillatory motion is proportional to the square root of the polarizability-to-mass ratio of the particle. This type of phenomenon is commonly observed in the fluorescence of cold atoms trapped within optical lattices \cite{PhysRevLett.69.49}.
As the polarizability-to-mass ratio is unique for different atomic or molecular species, a measurement of this characteristic signature frequency represents a new type of mass spectroscopy for 
these species, in their ground or excited states. Importantly, it does not require resonant optical fields and is therefore a general method that can potentially provide a complete elemental analysis of the measurement volume, in a single laser shot of typical duration of $\sim100-200$ ns. In this Letter we describe two ways to measure this unique oscillation frequency depending on the particle density at the measurement region: the first method uses light emission and is suitable for higher pressure mixtures while the second uses non-resonant laser ionization for the final detection of the species, which appears to be suitable for lower pressures. We note that what we proposed here is different to conventional Raman spectroscopy which probes the rotational and vibrational motion of molecules; instead, we probe the centre-of-mass motion of the particles confined by the deep optical lattice.

\textbf{Raman spectroscopy using light scattering: }In the high density case ($\gtrsim10^{22}$ m$^{-3}$)
, an all-optical detection scheme is proposed where the spontaneous Rayleigh emission from the trapped species is detected and spectrally measured. In the scheme, and the conditions considered in this Letter, a central peak due to the Dicke effect~\cite{PhysRevA.77.063409} appears in the spontaneous Rayleigh spectrum, while additional spectral sidebands are also present. The spectral location of these sidebands is proportional to the oscillation frequency of the particles within the potential well and is characteristic for each species. 

\begin{figure}[!h]
\vspace{-2mm}
\includegraphics[width=0.99\columnwidth]{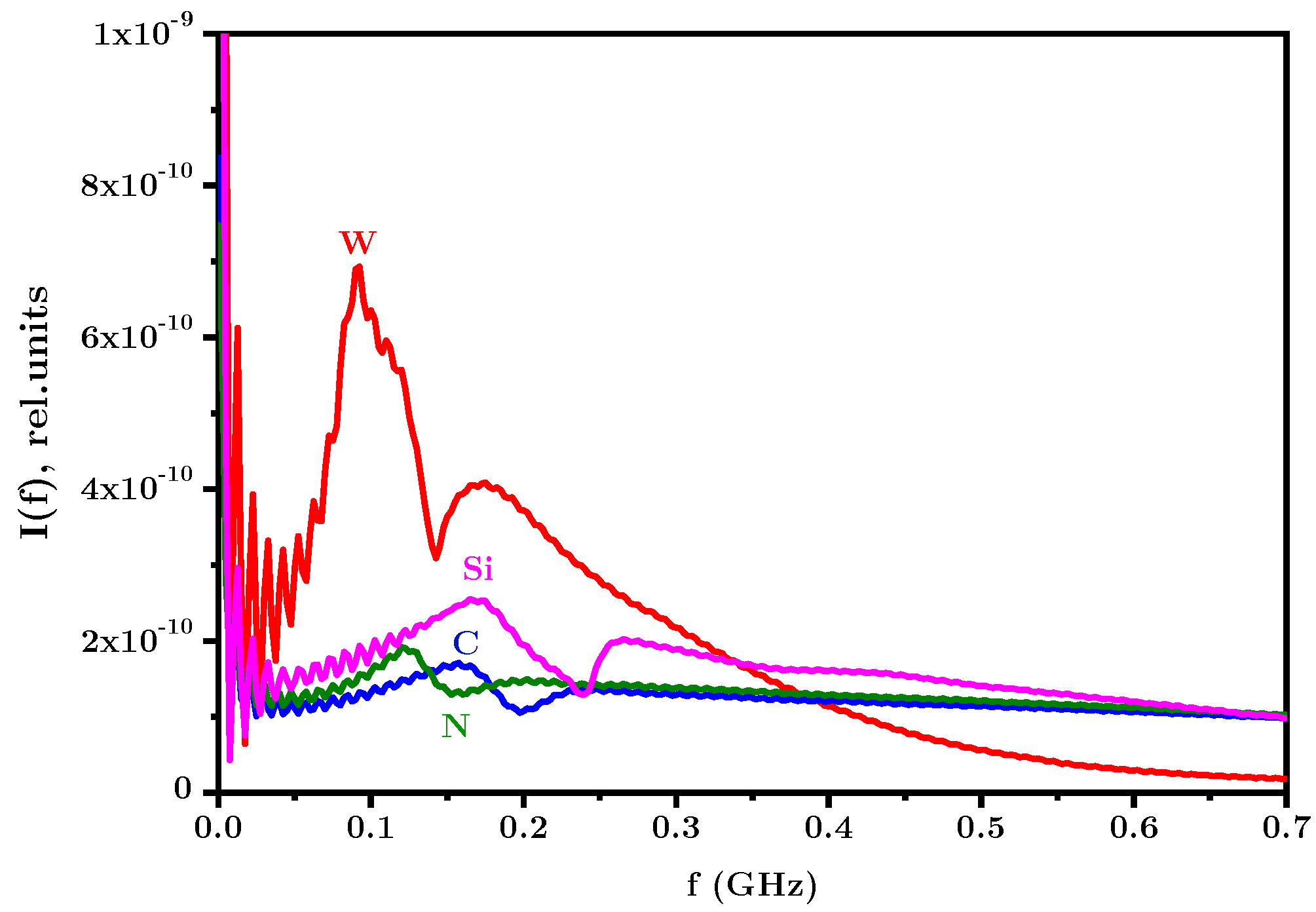}
\caption{\label{fig:Misha} 
Spectral profiles of Rayleigh scattered light of different atoms trapped in a pulsed optical lattice. In all cases the gas is assumed initially at a temperature of 300 K and a pressure of 1 Torr 
 for a $100$ ns pulse.}
\end{figure}

Calculations of the scattered light of both untrapped and trapped gas particles oscillating in the deep potential well of the optical lattice corresponds to the model described in Ref. \onlinecite{PhysRevA.77.063409}. These calculations are based on the solution of the one-dimensional unsteady kinetic Boltzmann equation, where the periodic dipole force of the lattice and the emission spectrum is calculated following the formalism provided in detail in Ref.~\onlinecite{Rautian_1967}. 
The calculations were carried out under the assumption of equal amplitudes of counter-propagating laser pulses of wavelength $\lambda$. Taking into account that in practice it is possible to control the temporal shape and duration of the laser pulse, the calculations were carried out for a super-Gaussian temporal pulse shape of $100$ ns pulse duration; qualitatively the behavior should be similar for shorter pulse durations. Calculations were performed for several gas species. For definiteness, it was assumed that the amplitude of each of the laser beams $I_{(a,1)}=I_{(a,2)}=10^{15}$ Wm$^{-2}$, a wavelength $\lambda$ of $532$  nm, and the temperature and gas pressure of T$_0=300$ K and  p$=1$ Torr, respectively. Figure \ref{fig:Misha} shows examples of calculations of optical spectra for $3$ different atomic components, W, C and N, obtained for an assumed laser pulse duration of $100$ ns.  The polarizabilities of atoms are $\alpha_i=4\pi\epsilon_0\tilde{\alpha}_i$, where $\tilde{\alpha}_i$ in $[10^{-30}$ m$^3]$, where the corresponding values $\tilde{\alpha}_i$ are taken from Ref. \onlinecite{lide2004crc}.

The particles trapped inside the pulsed potential well of the optical lattice perform periodic oscillations. These oscillations are manifested in the resulting signal spectrum by the presence of side peaks symmetrically located with respect to the center of the observed spectrum. A frequency shift of the sidebands can be estimated for particles at the bottom of the well where their oscillations are approximately harmonic. Considering the harmonic oscillatory motion in this area of each of the i-th species independently, the corresponding frequency shifts are determined by the simple formula~\cite{PhysRevA.77.063409}:

\begin{equation}
    f_i=\pm\frac{2}{\lambda}\sqrt{\frac{\alpha_i}{\epsilon_0cM_i}\sqrt{I_{(a,1)}I_{(a,2)}}}.
\end{equation}

Here, $M_i$, $\alpha_i$ are are the masses and polarizabilities of the i-th components of the gas mixture, respectively. 
It is assumed that the laser beams are exactly counter-propagating, resulting in the spatial period of the lattice (i.e. the spatial width of the potential well) $\lambda/2$. For identical intensities lasers where  $I_{(a,1)}=I_{(a,2)}=I$, we obtain a universal relation for various atomic components:

\begin{equation}
    \frac{|f_i|\lambda}{I^{1/2}}=\frac{2}{(\epsilon_0c)^{1/2}}\sqrt{\frac{\alpha_i}{M_i}}
\end{equation}

\noindent presented in Table 1. The estimates of the $f_i$ values are in good agreement with the results of numerical calculations shown in Fig.~\ref{fig:Misha}. Small differences are apparently associated with the time dependence of the laser intensity in numerical calculations and with the unharmonic nature of the oscillations in the realistic lattice potentials.

\begin{table}[h]
\centering
\begin{tabular}{@{}cccc@{}}
\toprule
 & \textbf{$\tilde{\alpha_i}$ {[}$10^{-30}$m$^3${]}} & \textbf{M$_i$ {[}kg{]}/$1.66\cdot 10^{-27}$} & \textbf{$\frac{f_i\lambda}{I^{1/2}}$} \\ \midrule
\textbf{Li} & $24.3$  & $6.941$   & $1.879\cdot10^4$  \\
\textbf{B}  & $30.3$  & $10.811$  & $5.3178\cdot10^3$ \\
\textbf{C}  & $1.76$  & $12.011$  & $3.845\cdot10^3$  \\
\textbf{Na} & $24.08$ & $22.9877$ & $1.028\cdot10^4$  \\
\textbf{Al} & $6.8$   & $26.981$  & $5.0427\cdot10^3$ \\
\textbf{Si} & $5.38$  & $28.085$  & $4.396\cdot10^3$  \\
\textbf{Fe} & $8.4$   & $35.847$  & $4.862\cdot10^3$  \\
\textbf{W}  & $11.1$  & $183.85$  & $2.468\cdot10^3$  \\
\textbf{Au} & $5.8$   & $196.966$ & $1.7237\cdot10^3$ \\
\textbf{N}  & $1.1$   & $14.0$    & $2.8156\cdot10^3$ \\ \bottomrule
\end{tabular}
\caption{\textbf{Frequency $f_i$ of each species $i$ measured in [Hz], wavelength $\lambda$ in [nm] and laser intensity $I$ in [$\frac{W}{m^2}$]}}
\label{tab:my-table}
\end{table}

\textbf{Raman spectroscopy using ionization detection: } At lower densities, it is anticipated that detecting the few scattered photons through spontaneous light scattering can be challenging. To address this, we present a complimentary detection scheme, suitable at low species densities, which utilizes the coherent oscillation and subsequent non-resonant ionization of species within deep optical lattices. The high efficiency detection of the resulting charged species occurs as they are extracted towards a microchannel plate (MCP) detector which is at a high negative potential. The different species will oscillate at different frequencies within these deep optical potentials, while a fraction are ionized when reaching the bottom of the well. Recording the arrivals of charged species at the detector and obtaining the corresponding power spectrum, allows for the precise, non-resonant detection of all species in the measurement volume and the differentiation of even different isotopes within it.  

\begin{figure}
\includegraphics[width=0.99\columnwidth] {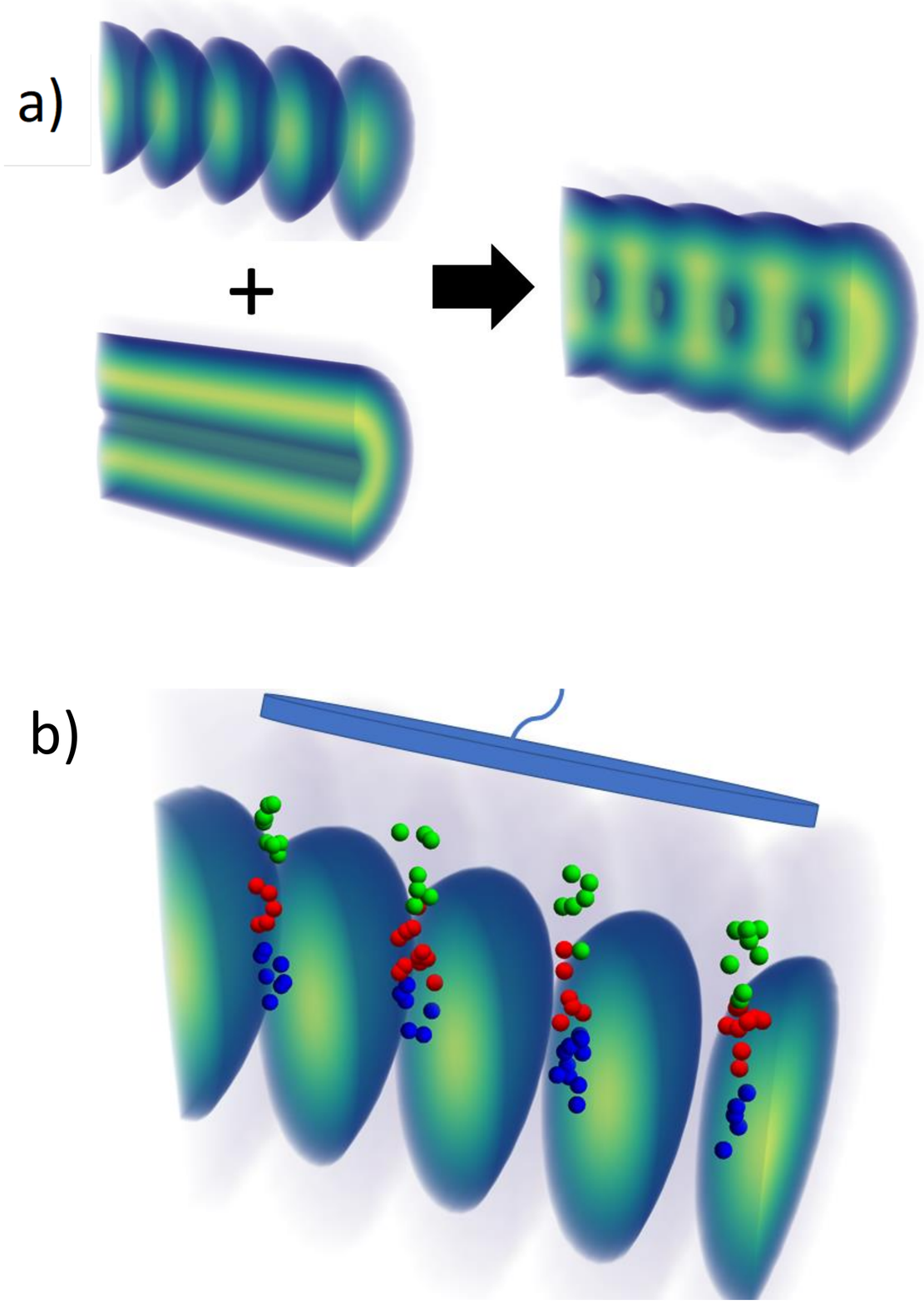}
\caption{\label{fig:donut}a) Combination of optical lattice with a hollow cylindrical (donut) beam to ionize most of the species leaving only a well localized neutral gas region in the periodic dark regions along the center where the light intensity is low b) Following ionisation, the hollow cylindrical beam is turned off and the lattice intensity is reduced. The neutral particles then oscillate in the lattice. During their oscillation a few of the different neutral particles are ionized (blue, red and green balls) and are collected by an MCP plate (blue plate above lattice), which measures the periodic modulation in total charge. The time that the species are ionized and therefore their detection on the MCP depends on their oscillation frequency within the lattice}
\end{figure}

The key to obtaining narrow spectral features is to remove most 
species from within the lattice, leaving only a well-localized density that will coherently oscillate in the nearly harmonic optical potential of the lattice. This can be achieved by initially using a high intensity ionizing lattice pulse, whose phase is shifted by $90$ degrees with respect to the lattice used to perform the spectroscopy.  In addition to this lattice pulse, it is also desirable to use a hollow cylindrical beam to provide radial ionization so that only neutral atoms are left within the center of the transport lattice (Fig.~\ref{fig:donut}a). After this step each species undergoes oscillatory motion at different frequencies for an optical pulse duration on the order of $50$ ns. As they oscillate in the lattice, a small fraction will experience periodic ionization in the regions of high intensity, due to the high optical fields utilized, leading to an oscillatory production of ions. Detecting this periodic arrival of ions, directly relates to the oscillatory motion of the neutral species in the lattice and thus to their unique polarizability-to-mass ratio.


\begin{figure}
\includegraphics[width=0.99\columnwidth] {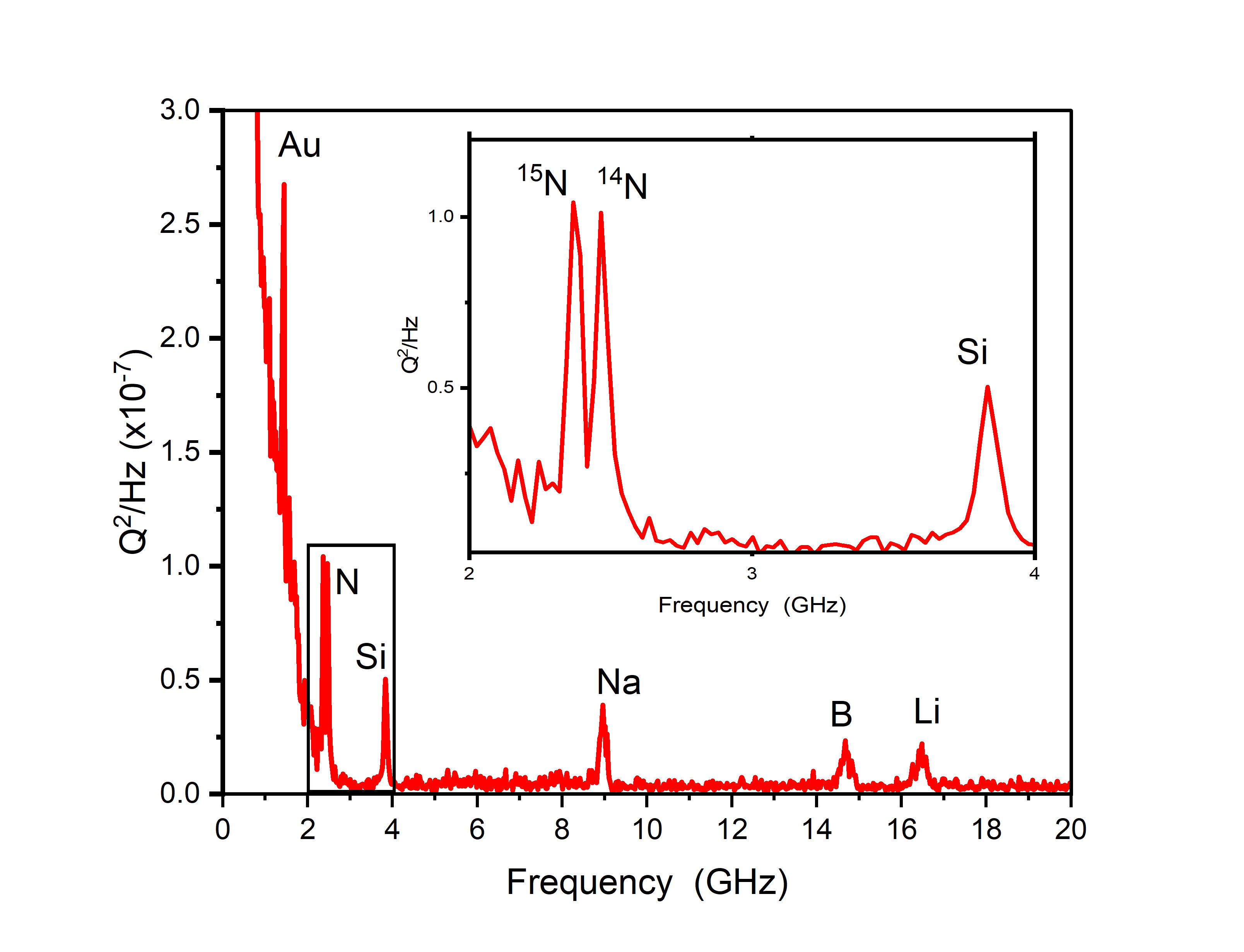}
\caption{\label{fig:powspec} The motional spectrum as derived from the power spectral density of the time series of charges produced by an equal mixture of atoms (Au, N, Si, Na, B and Li) by ionization oscillating within the pulsed optical lattice. The inset graph is a magnified view of spectral features indicating that nitrogen 15 and 14 can be spectrally isolated.}
\end{figure}

To calculate the dynamics of the particles in this lattice field we use a Monte Carlo approach where $80,000$ atoms are placed within the lattice region. We only need to calculate motion within a single lattice period of $\lambda/4$ since similar motion is replicated in each well of the lattice. Each atom is then allowed to move under the action of the optical dipole force of the lattice, in $3$-D, as a function of time.  To observe the oscillatory dynamics at low pressure, where it is not feasible to extract a strong scattering signal, we use a higher intensity lattice so that a small fraction of the total number of atoms is ionized during the transport lattice. This leads to the multiphoton ionization of atoms during their oscillatory dynamics. Although the ion fraction is small, the ions can be recorded with near unit efficiency and thus only a small ionization rate suffices. We calculate the ionization probability as a function of time using the above threshold ionization formula of Ammosov, Delone, and Krainov~\cite{10002999876}. The ionization probability at each time step is compared with a random probability distribution between $0$ and $1$ and, if it is less than this value, the atom is ionized.  The number of charges is stored at each time step and the power spectrum is calculated to determine the oscillatory spectra. These give an estimate of the power spectral density per pulse and are shown in Fig.~\ref{fig:powspec} for a gas consisting of equal components of Au, Si, Na and Li atoms in the measurement region. These species have very different polarizability to mass ratios. A lattice intensity of $3.5\cdot10^{17}$ W m$^{-2}$ was used. Also shown in the inset of Fig.~\ref{fig:powspec}, is the ability to resolve even isotopes of $^{15}$N and $^{14}$N atoms which are of importance for some applications. Note that for higher polarizability-to-mass ratio species the spectral widths increase. This is due to the larger dispersion of the atoms as they oscillate over more periods than the lower polarizability-to-mass ratio species.

With an ionized gas, a quasi-neutral plasma is formed with an ion density $n_+$ equal to the electron density $n_e$. For this diagnostic, it is necessary that the emerging ions freely leave the region of their generation. This occurs if the size of the Debye layer $r_D=(\frac{\epsilon_0k_BT_e}{e^2n_+})^{1/2}$, where $\epsilon_0$ is the vacuum permittivity, $k_B$ is Boltzmann's constant, $T_e$ and $e$ the electron temperature and charge, respectively, exceeds the characteristic size of the generation region. For this estimate, it is reasonable to assume that the characteristic size of the ionization region is of the order of the radius of the optical lattice, $R_L$. The condition for free (non-ambipolar) expansion of the formed ions is $r_D>R_L$. Thus, the density of charged particles of the resulting plasma is limited: $n_+<\frac{\epsilon_0k_BT_e}{e^2R^2_L}$. For example, for $R_L=100$ $\mu$m and an electron temperature for the generated electrons $T_e\sim1$ eV, the density of charged particles of the resulting plasma should not exceed $n_+<5.5\cdot10^{15}$ m$^{-3}$. This estimate forms an upper limit for this technique; it is anticipated that in an experiment the ideal density should be one order of magnitude less. 

The oscillatory Raman mass-spectrometry considered here, based on the induced oscillations of the centre-of-mass motion of atoms and molecules within deep periodic optical potentials, offers a novel and very general approach to quantitative species detection within gases. This technique can be interfaced with already established mass spectrometry techniques such as SIMS, MALDI or the Orbitrap, enhancing their detection performance and potentially overcoming current limitations. 

Our calculations show that this technique can be applied over a wide range of densities and species and that even isotopic species can be well distinguished using this technique. Just as the charge-to-mass ratio has been successfully employed for the detection of species within mass spectrometers, the dispersion based on the polarizability-to-mass ratio as demonstrated here offers potentially even greater selectivity: the polarizability, which is dependent on both the species and the electronic state, can potentially give information on both the ground state and excited states fractions of different species. This is possible since the oscillatory dynamics is faster than typical excited state lifetimes. Finally, we have focused on two methods to measure the polarizability-to-mass, one consisting of using light spontaneous scattered from the trapped particles within the lattice. It is envisioned that both methods can be incorporated in current state-of-the-art mass spectrometry instruments and work in a complimentary way. Although here we are only considering the spontaneous light scattering case, it is envisioned that a coherent scattering equivalent can be realized, where a probe beam, incident at the Bragg angle~\cite{Bragg428} on the resulting optical lattice is coherently scattered from it, giving rise to a fourth signal beam. A similar scheme has been successfully demonstrated for single shot gas~\cite{Gerakis:13, doi:10.1063/1.4959778} and nanoparticle diagnostics~\cite{PhysRevApplied.9.014031}.

\begin{acknowledgments}
AG would like to acknowledge Dr. Tom Wirtz of LIST for fruitful discussions and feedback on the manuscript. 
\end{acknowledgments}

\section*{Data Availability Statement}

The data that support the findings of this study are available from the corresponding author upon reasonable request

\appendix

\nocite{*}
\bibliography{aipsamp}

\end{document}